\documentclass[a4paper,11pt]{article}
\pdfoutput=1 

\usepackage{jheppub} 
\usepackage[T1]{fontenc} 
\usepackage[english]{babel}
\usepackage{graphicx}
\usepackage{epstopdf}
\usepackage{epsfig}
\newcommand{\ep}{\varepsilon}

\title{Counting the number of master integrals for sunrise diagrams via the Mellin--Barnes representation}

\author{Mikhail Yu.\ Kalmykov}
\author{Bernd A.~Kniehl}
\affiliation{II. Institut f\"ur Theoretische Physik, Universit\"at Hamburg,
             Luruper Chaussee 149, 22761 Hamburg, Germany}
\emailAdd{kalmykov.mikhail@gmail.com}
\emailAdd{kniehl@desy.de}

\abstract{
A number of irreducible
master integrals for $L$-loop sunrise and bubble Feynman diagrams
with generic values of masses and external momenta are explicitly evaluated
via the Mellin--Barnes representation.
}

\begin{document} 
\maketitle
\flushbottom

\section{Introduction}

\begin{figure}[th]
\centering 
\includegraphics[width=60mm,height=80mm]{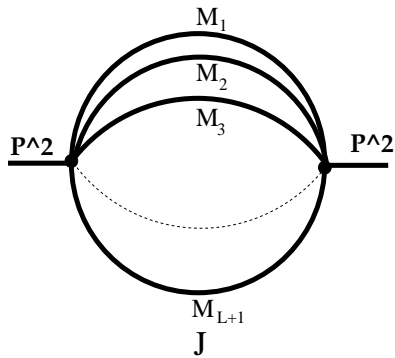}
\caption{\label{sunrise}
$L$-loop sunrise diagram.}
\end{figure}

The sunrise or watermelon diagram (see Fig.~\ref{sunrise}) 
is one of the simplest Feynman diagrams which have been studied 
by the physics community as well as by mathematicians over the past fifty years 
\cite{Ponzano,sunrise:picard,Mendels,sunrise:DE,sunrise:2loop,sunrise:modular,%
  sunrise:integral,sunrise:tarasov,groote,borwein,sunrise:motivies,sunrise,%
KT:2016}.
This diagram has a few different representations.  
Within dimensional regularization \cite{dimreg} in momentum space, it is defined as 
\begin{eqnarray}
&& \hspace{-5mm}
J(\vec{M_j^2};\vec{\alpha_j};p^2)
= 
\int 
\prod_{j=1}^L \frac{d^n k_j}{[k_j^2\!-\!M_j^2]^{\alpha_j} } 
\times
\frac{1}{[(p\!-\!k_1\!-\cdots\!-\!k_L)^2\!-\!M_{L+1}^2]^{\alpha_{L+1}}} \;, 
\nonumber \\ && 
\label{sunset}
\end{eqnarray}
where $\alpha_j$ are positive integers, $M_j^2$ and $p^2$ 
are some (in general, complex) parameters, and $n$ 
is a (in general, non-integer) parameter of dimensional regularization.
The parametric representation of this diagram has the following form:\footnote{
It could also be rewritten in projective space \cite{Ponzano,sunrise:picard}.}
\begin{eqnarray}
J 
\sim  
\Gamma \left(\alpha - \frac{n}{2}L \right)
\int_0^\infty \prod_{i=1}^{L+1} 
       d x_i \frac{x_i^{\alpha_i-1}}
                  {\Gamma\left(\alpha_i\right)} 
              \delta \left(1-\sum_i x_i \right)
\frac{F^{\frac{n}{2}L-\alpha}}
     {U(x)^{\frac{n}{2}(L+1)-\alpha}} \;,
\label{sunrise:parametric}
\end{eqnarray}
where $F$ and $U$ are Symanzik polynomials \cite{BW:2010} and 
$\alpha = \sum_{j=1}^{L+1} \alpha_j$.
Using the coordinate representation of the Feynman propagator
and integrating over the angle, it is easy to get \cite{Mendels}
a one-fold integral representation of this diagram 
(for details, see Refs.~\cite{groote,borwein} and Appendix~\ref{appendixB}). 
Applying the algorithm of Ref.~\cite{BD}, a Mellin--Barnes integral representation 
\cite{barnes} for the sunrise diagram can be deduced \cite{sunrise:2loop} 
(see Eq.~(\ref{mellin-barnes:sunset})).

The aim of the present paper is to extend the approach described in Refs.~\cite{counting,KK2012} 
to the multivariable case with reducible monodromy.
As an illustration, the number of irreducible master integrals of the $L$-loop
sunrise diagram is evaluated. 
In Appendix \ref{appendixA}, the generalized sunrise diagram is considered.
The application of our analysis to integrals including products of Bessel functions is discussed in Appendix~\ref{appendixB}.

\section{Mellin--Barnes integral versus Horn hypergeometric function}
\label{MB}
Let us consider the function $\Phi$ defined through a $K$-fold Mellin--Barnes integral,
\begin{eqnarray}
&& 
\Phi(A,B;\{C_k\};\{z_j\})
= 
\int 
\prod_{j=1}^{K}
dt_j 
\Gamma(-t_j) \Gamma\left(C_j\!-\!t_j \right)
z_j^{t_j}
\frac{
      \Gamma\left(A \!+\! t \right)
     }
     {\Gamma\left(B \!-\! t \right)
     } \;,
\label{Phi:MB}
\end{eqnarray}
where 
$
t = \sum_{j=1}^K t_j \;
$
and 
$\{z_a\} = (z_1, z_2, \cdots, z_K) \;$.
This function depends on $K+2$ discrete parameters $A$, $B$, and $\{C_j\}=(C_1,\cdots, C_K)$
and $K$ variables $z_1,\cdots,z_K$.
Let us briefly recall some basic steps of the differential-reduction 
procedure \cite{algorithm} applied to the Mellin--Barnes integral.
The differential contiguous relations for the function $\Phi$
follow directly from the Mellin--Barnes representation 
and have the following form:
\begin{subequations}
\begin{eqnarray}
\Phi(A,B;\{ \}, C_j+1, \{ \}; \{z_k\})  & = & 
(C_j \!-\! \theta_j) \Phi(A,B;\{ \}, C_j, \{ \}; \{z_k\})  
\;, 
\label{Phi:C:contiguous}
\\ 
\Phi(A+1,B; \{C_k\}; \{z_k\})  & = & 
(A \!+\! \sum_{j=1}^{K} \theta_j) \Phi(A,B;\{C_k \}; \{z_k\})  
\;, 
\label{Phi:A:contiguous}
\\ 
\Phi(A,B-1; \{C_k \}; \{z_k\})  & = & 
(B \!-\! 1 \!-\! \sum_{j=1}^{K} \theta_j)
\Phi(A; B; \{C_k \}; \{z_k\})  \;, 
\label{Phi:B:contiguous}
\end{eqnarray}
\label{Phi:contiguous}
\end{subequations}
where
$$
\theta_j = z_j \frac{d}{d z_j} \;, \quad j=1,\cdots,K.
$$
We denote the set of differential operators on the r.h.s.\ of Eq.~(\ref{Phi:contiguous}) 
as ${\bf B}_{C_j,A,B}^{+}$. 
A linear system of partial differential equations (PDEs) for 
the function $\Phi$ can be derived in two steps.
In the first step, we define the polynomials $P$ and $Q$ as 
$$
\frac{P_j}{Q_j} =  \frac{\phi(t_j+1)}{\phi(t_j)} \;. 
$$
In the second step, we set up the corresponding system of PDEs,
$$
L_j:
\left( 
\left. Q_j \right|_{t_j \to \theta_j} \frac{1}{z_j} \Phi 
=  
\left. P_j \right|_{t_j \to \theta_j}  \Phi 
\right) \;,
$$
where, for simplicity, we have introduced the short-hand notation: $\Phi \equiv \Phi(A,B; \{C_k\}; \{z_k\})$.
For the function under consideration, we have
\begin{subequations}
\begin{eqnarray}
&& 
\frac{P^{\Phi}_j}{Q^{\Phi}_j}  =  
- \frac{\left(A+t\right)\left(1-B+t \right)}
       {(1-C_j+t_j)(1+t_j)} 
\Rightarrow  
\\ && 
L_j^{\Phi}: 
\left( \theta_j \!-\! C_j \right) \theta_j \Phi 
= 
-  
z_j 
\left(\sum_{j=1}^{K} \theta_j \!+\! A\right)
\left(\sum_{j=1}^{K} \theta_j \!+\! (1-B)\right)
\Phi \;, 
\quad j=1,\cdots,K. 
\label{sunrise:PDE}
\end{eqnarray}
\end{subequations}
To get the full system of PDEs for the function $\Phi$,
a prolongation procedure should be applied \cite{Lauricella,Cartan}, which
consists in applying new derivatives to the system of PDEs,
so that the system of PDEs in Eq.~(\ref{sunrise:PDE}) can be written in a Pfaffian form:\footnote{%
The condition of complete integrability 
is valid.}
\begin{eqnarray}
d \vec{\phi} = \Omega \vec{\phi} \;,
\label{Pfaffian}
\end{eqnarray}
where 
the matrix $\Omega$ only depends on the values of the parameters 
and the singular locus of the system of PDEs, 
and the vector function $\vec{\phi}$ is defined as 
$(\Phi, \theta_i \Phi, \theta_{ij} \Phi, \cdots, \theta_{j_1,j_2,\cdots,j_m} \Phi)$.
The rank $r$ of the matrix $\Omega$ at the point $z_0$ (in our case, $z_0=\vec{z}_0=0$)  in Eq.~(\ref{Pfaffian})
is equal to the number of independent solutions of the full system of PDEs. 

According to the algorithm described in Ref.~\cite{algorithm},
the differential operators $b_j^{-}$ inverse to the operators
defined by Eq.~(\ref{Phi:contiguous}) can be constructed so that
\begin{equation}
b_j^{-} {\bf B}_j^{+} \Phi(A,B;\{C_j\};\{z_j\}) 
=
\Phi(A,B;\{C_j\};\{z_j\})  \;, \quad j = 1, \cdots, K \;.
\end{equation}
The operators $b_a^{-}$ are defined\footnote{Due to the differential relation 
$$
\theta_p \Phi(A,B;\{C_j\};\{z_j\}) = -z_p \Phi\left(1+A,B-1;C_p-1,\{C_j\};\{z_j\}\right)
\equiv - z_p B_A^{+} B_B^{+} b_C^{-} \Phi(A,B;\{C_j\};\{z_j\}), 
$$
not all operators $b_c^{-}$ are independent.} 
modulo the full system of PDEs.
The differential reduction has the form of a product of several operators
$b_i^{-}$ and $B_j^{+}$. 
In symbolic form, this can be written as
\begin{eqnarray}
\Phi(\vec{I}+\{A,B,\{C_j\}\};\{z_j\}) 
=
\left[
Q_0
\!+\! \sum_{j=1}^K Q_j \theta_j
\!+\! \sum_{\substack{i,j=1 \\ i<j}}^K Q_{ij} \theta_{ij}
+ 
\cdots 
\right] 
\Phi(A,B,\{C_j\};\{z_j\}) \;,
\label{reduction}
\end{eqnarray}
where $\vec{I}$ is a set of integers, $\{A,B,\{C_j\} \}$ is a set of parameters,
and $Q_i$ are some rational functions of $\{z_i\}$ and $A,B,\{C_j\}$.

%
%

The fundamental system of solutions of Eq.~(\ref{sunrise:PDE}) is expressible in terms of the 
Lauricella \cite{Lauricella} function $F_C^{(K)}$ of $K$ variables and include the following functions
(see, for example, Eq.~(13) in Ref.~\cite{Lauricella} or Eq.~(19) in Ref.~\cite{vilenkin}): 
\begin{eqnarray}
&& 
F_C^{(K)}(A,1-B;\{1\!-\!C_j\};\{-z_j\}) \;, 
\nonumber \\ &&
(-z_j)^{C_j}  
\times F_C(A\!+\!C_j,1\!-\!B\!+\!C_j;1\!-\!C_1, \cdots, 1\!-\!C_{j-1}, 1\!+\!C_j, 1\!-\!C_k;\{-z_j\}) \;,
\nonumber \\ && \hspace{5mm}
j=1, \cdots, K \;,  
\nonumber \\ &&
(-z_{j_1})^{C_{j_1}} 
(-z_{j_2})^{C_{j_2}} 
               \times
               F_C^{(K)}(A\!+\!C_{j_1}\!+\!C_{j_2},
                    1\!-\!B\!+\!C_{j_1}\!+\!C_{j_2};\{1\!-\!C_k\},1\!+\!C_{j_1},1\!+\!C_{j_2};\{-z_j\}) \;, 
\nonumber \\ && \hspace{5mm}
j_1,j_2=1, \cdots, K \;,  
\nonumber \\ &&
\cdots 
\nonumber \\ &&
\prod_{j_a=1}^K
   (-z_{j_a})^{C_{j_a}} 
            \times
          F_C^{(K)}(A\!+\!\sum_{j=1}^K C_j,1\!-\!B\!+\!\sum_{j=1}^K C_j;\{1+C_{j}\};\{-z_j\}) \;, 
\label{Phi-FundamentalSolution}
\end{eqnarray}
where the Lauricella function $F_C^{(K)}$ is defined as 
\begin{eqnarray}
F_C^{(K)}(a,b;\{c_j\};\{z_j\}) = 
\sum_{j_1,\cdots,j_k=0}^\infty (a)_{j_1+\cdots+j_K} (b)_{j_1+\cdots+j_K} 
                             \prod_{p=1}^K \frac{z_p^{j_p}}{j_p! (c_p)_{j_p}} \;,
\label{FC}
\end{eqnarray}
with $(a)_k = \Gamma(a+k)/\Gamma(a)$ being the Pochhammer symbol.
Any particular solution of the system defined by Eq.~(\ref{sunrise:PDE}) about the points $z_i=0$ for a generic 
set of parameters is a linear combination of solutions defined by Eq.~(\ref{Phi-FundamentalSolution})
with undetermined coefficients.
To fix these coefficients, it is necessary to evaluate the Mellin--Barnes integral as a power series 
solution \cite{beukers}.
In particular, under the condition that the monodromy is irreducible (see Eq.~(\ref{exceptional:phi})),  
the following representation is valid (see Ref.~\cite{BD} for details):\footnote{The interrelations between Mellin--Barnes integrals 
and multiple residues were discussed in Ref.~\cite{MB:residues}.}
\begin{eqnarray}
&&
\Phi(A,B,\{C_j\};\{z_j\}) =
\frac{\Gamma\left(A\right) }
     {\Gamma\left(B\right) }
\prod_{j=1}^K \Gamma\left( C_j \right)  \times F_C^{(K)}(A,1-B;\{1\!-\!C_j\};\{-z_j\})
\nonumber \\ &&
+
\sum_{j=1}^K   (-z_j)^{C_j} \frac{\Gamma\left(A\!+\!C_j\right) \Gamma\left( -C_j \right)  }
                                 {\Gamma\left(B\!-\!C_j\right) }
               \prod_{\substack{a=1 \\ a \neq j}}^K  \Gamma \left( C_a \right)  
\nonumber \\ && \hspace{5mm}
               \times
               F_C \left(A\!+\!C_j,1\!-\!B\!+\!C_j;1\!-\!C_1, \cdots, 1\!-\!C_{j-1}, 1\!+\!C_j, 1\!-\!C_k;\{-z_j\}
                   \right)
\nonumber \\ &&
+
\sum_{j_1,j_2=1}^K
   (-z_{j_1})^{C_{j_1}} \Gamma\left( -C_{j_1} \right)  
\times
   (-z_{j_2})^{C_{j_2}} \Gamma\left( -C_{j_2} \right)
\times 
\frac{\Gamma\left(A\!+\!C_{j_1}\!+\!C_{j_2}\right)} 
     {\Gamma\left(B\!-\!C_{j_1}\!-\!C_{j_2}\right)}
\nonumber \\ && \hspace{5mm}
               \times
                \prod_{\substack{a=1 \\ a \neq j_1,j_2}}^K  \Gamma \left( C_a \right)  
               \times
               F_C^{(K)}\left(A\!+\!C_{j_1}\!+\!C_{j_2},
                    1\!-\!B\!+\!C_{j_1}\!+\!C_{j_2};\{1\!-\!C_k\},1\!+\!C_{j_1},1\!+\!C_{j_2};\{-z_j\}\right)
\nonumber \\ &&
+
\cdots
\nonumber \\ &&
+
\sum_{b=1}^K 
\frac{\Gamma\left(C_{j_b} \right) }
     {(-z_{j_b})^{C_{j_b}} \Gamma\left( -C_{j_b} \right)} 
\times 
    \frac{\Gamma\left(A \!+\! \sum_{j=1}^K C_{j}\!-\!C_b \right)} 
         {\Gamma\left(B \!-\! \sum_{j=1}^K C_{j}\!-\!C_b\right) }
\times 
\prod_{a=1}^{K} (-z_{j_a})^{C_{j_a}} \Gamma\left( -C_{j_a} \right)
\nonumber 
\\ && \hspace{5mm}
           \times
F_C^{(K)}(A\!+\!\sum_{j=1}^K C_j \!-\! C_b ,1\!-\!B\!+\!\sum_{j=1}^K C_j \!-\! C_n;1-C_b,\{1+C_{j}\};\{-z_j\})  
\nonumber \\ &&
+
\prod_{j_a=1}^K
   (-z_{j_a})^{C_{j_a}} \Gamma\left( -C_{j_a} \right) 
                     \frac{\Gamma\left(A \!+\! \sum_{j=1}^K C_{j} \right)} 
                          {\Gamma\left(B \!-\! \sum_{j=1}^K C_{j}\right) }
\nonumber \\ && \hspace{5mm}
            \times
          F_C^{(K)}(A\!+\!\sum_{j=1}^K C_j,1\!-\!B\!+\!\sum_{j=1}^K C_j;\{1+C_{j}\};\{-z_j\}) \;. 
\label{Phi<->FC}
\end{eqnarray}
We wish to point out that, for the sunrise diagram, 
the last term in Eq.~(\ref{Phi<->FC}) is proportional to $1/\Gamma(0)$ and so equal 
to zero.\footnote{Another example of such a cancellation was presented in Ref.~\cite{MKL2009}.}

The holonomic\footnote{We adopt the following definition of holonomic function \cite{algorithm}: 
a function is called holonomic if it satisfies a system of linear differential equations 
with polynomial coefficients whose solutions form a finite-dimensional vector space. 
} 
rank\footnote{The dimension of the space of solutions of a system of PDEs near some 
generic point is called holonomic rank.} of the system of PDEs in Eq.~(\ref{sunrise:PDE})
for a generic set of parameters $\{A,B,C_j\}$ is equal to $2^{K}$ \cite{Lauricella} 
(see also Refs.~\cite{FC:1,FC:2,FC:3,FC:hyperdire}).  
However, if the parameters $A,B,\{C_i\}$ satisfy certain linear relations, then
additional differential operators are generated, so that Puiseux-type solutions appear.
The main questions are how to find such linear relations between the parameters
and how to define a minimal set of the additional PDEs.
Our approach \cite{counting,KK2012} to these problems is based on studying the inverse differential operators: 
the exceptional case of parameters, where the dimension of the solution space is reduced,
corresponds to the condition that the denominators of the functions $Q_i$ entering
Eq.~(\ref{reduction}) are equal to zero for arbitrary values of $z_i$ \cite{HYPERDIRE}.
The same recipe works in its application to Mellin--Barnes integrals 
\cite{beukers}, which can be treated as a particular case of the
Gelfand--Kapranov--Zelevinsky (GKZ) hypergeometric system \cite{GKZ}.
However, the inverse differential operators have a very complicated structure 
(see, for example, Ref.~\cite{FC:hyperdire}), which gives rise to technical
problems in the analysis of the number of independent PDEs.

Fortunately, there is a simpler way 
\cite{beukers,GKZ:monodromy,monodromy,sadykov})
to find the conditions of reducibility and to define the dimension of the
(ir)reducible subspace of solutions.
In our case, the system of PDEs is irreducible if 
(for details, see Section~\ref{dimension})
\begin{subequations}
\begin{eqnarray}
F_C & : &
\Biggl \{
a \notin \mathbb{Z}, ~~
b \notin \mathbb{Z}, ~~
a-\sum_{s_1} c_p \notin \mathbb{Z}, ~~
b-\sum_{s_2} c_p \notin \mathbb{Z}  
\Biggr \} \;,
\label{exceptional:fc}
\\
\Phi &:&
\Biggl \{
A \notin \mathbb{Z}, ~~
B \notin \mathbb{Z}, ~~
A+\sum_{S_1} C_p \notin \mathbb{Z}, ~~
B-\sum_{S_2} C_p \notin \mathbb{Z}  
\Biggr \} \;, 
\label{exceptional:phi}
\end{eqnarray}
\label{exceptional}
\end{subequations}
\noindent
where $s_i$ and $S_j$ are any subsets of $\{c_i\}$ and $\{C_j\}$, respectively.
Eq.~(\ref{exceptional:fc}) 
corresponds to the set of exceptional values of the parameters
of the hypergeometric function $F_C$,\footnote{This set of parameters complies
with the condition that the monodromy group of the Lauricella function $F_C$ is reducible \cite{FC:1}.
For recent results on the evaluation of the monodromy of the GKZ system, see
Ref.~\cite{GKZ:monodromy}.}
and Eq.~(\ref{exceptional:phi}) defines the exceptional set of parameters for
the function $\Phi$.
The conditions defined by Eq.~(\ref{exceptional:phi}) are invariant with respect to a linear 
change of variables $\{t_j\} \to \{A_1 \pm \sum_{A \in \{1, \cdots, p \}} t_A \}$ 
in the Mellin--Barnes integral in Eq.~(\ref{Phi:MB}).

In a similar manner, 
we can consider the function $\Psi$ defined as a $K$-fold Mellin--Barnes integral,
\begin{eqnarray}
&& 
\Psi(A,D;\{C_k\};z_1,\cdots,z_K)
= 
\int 
\prod_{j=1}^{K}
dt_j 
\Gamma(-t_j) \Gamma\left(C_j\!-\!t_j \right)
z_j^{t_j}
\Gamma\left(A \! +\! t \right)
\Gamma\left(D \!+ \! t \right)
\;.
\label{Psi:MB}
\end{eqnarray}
In this case, we have
\begin{subequations}
\begin{eqnarray}
&& 
\frac{P^{\Psi}_j}{Q^{\Psi}_j}  =  
\frac{\left(A+t \right)\left(D+t \right)}
     {(1-C_j+t_j)(1+t_j)} 
\Rightarrow  
\\ && 
L_j^{\Psi}: 
\left( \theta_j \!-\! C_j \right) \theta_j \Psi 
= 
z_j 
\left(\sum_{j=1}^{K} \theta_j \!+\! A\right)
\left(\sum_{j=1}^{K} \theta_j \!+\! D \right)
\Psi \;, 
\quad j=1,\cdots,K. 
\label{bubble:PDE}
\end{eqnarray}
\end{subequations}
The system of PDEs for the function $\Psi$ is irreducible if 
\begin{eqnarray}
\Psi &:&
\Biggl \{
A \notin \mathbb{Z}, ~~
D \notin \mathbb{Z}, ~~
A+\sum_{S_1} C_p \notin \mathbb{Z}, ~~
D+\sum_{S_2} C_p \notin \mathbb{Z} 
\Biggr \} \;,
\label{exceptional:psi}
\end{eqnarray}
where $S_1$ and $S_2$ are any subsets of $\{C_j\}$.
The solution of Eq.~(\ref{bubble:PDE}) for a generic set of parameters 
about the points $z_i=0$ can be written as linear combination of Lauricella 
functions $F_C^{(K)}$ of $K$ variables.
For completeness, we present it here:
\begin{eqnarray}
&&
\Psi(A,D,\{C_j\};\{z_j\}) =
\Gamma\left(A\right)\Gamma\left(D\right)
\prod_{j=1}^K \Gamma\left( C_j \right)  \times F_C^{(K)}(A,D;\{1\!-\!C_j\};\{z_j\})
\nonumber \\ &&
+
\sum_{j=1}^K   (z_j)^{C_j} \Gamma\left(A\!+\!C_j\right) 
                           \Gamma\left(D\!+\!C_j\right) 
                           \Gamma\left( -C_j \right)                           
               \prod_{\substack{a=1 \\ a \neq j}}^K  \Gamma \left( C_a \right)  
\nonumber \\ && \hspace{5mm}
               \times
               F_C^{(K)}(A\!+\!C_j,D\!+\!C_j;1\!-\!C_1, \cdots, 1\!-\!C_{j-1}, 1\!+\!C_j, 1\!-\!C_K;\{z_j\})
\nonumber \\ &&
+
\sum_{j_1,j_2=1}^K
   (z_{j_1})^{C_{j_1}} \Gamma\left( -C_{j_1} \right)  
\times
   (z_{j_2})^{C_{j_2}} \Gamma\left( -C_{j_2} \right)
\times 
\Gamma\left(A\!+\!C_{j_1}\!+\!C_{j_2}\right) 
\times
\Gamma\left(D\!+\!C_{j_1}\!+\!C_{j_2}\right)
\nonumber \\ && \hspace{5mm}
               \times
                \prod_{\substack{a=1 \\ a \neq j_1,j_2}}^K  \Gamma \left( C_a \right)  
               \times
               F_C^{(K)}(A\!+\!C_{j_1}\!+\!C_{j_2},
                         D\!+\!C_{j_1}\!+\!C_{j_2};\{1\!-\!C_k\},1\!+\!C_{j_1},1\!+\!C_{j_2};\{z_j\})
\nonumber \\ &&
+
\cdots
\nonumber \\ &&
+
\sum_{b=1}^K 
\frac{\Gamma\left(C_{j_b} \right) }
     {(z_{j_b})^{C_{j_b}} \Gamma\left( -C_{j_b} \right)} 
\times 
    \Gamma\left(A \!+\! \sum_{j=1}^K C_{j}\!-\!C_b \right)
\times 
    \Gamma\left(D \!+\! \sum_{j=1}^K C_{j}\!-\!C_b \right)
\nonumber \\ && \hspace{2mm}
\times 
\prod_{a=1}^{K} (z_{j_a})^{C_{j_a}} \Gamma\left( -C_{j_a} \right)
           \times
F_C^{(K)}(A\!+\!\sum_{j=1}^K C_j \!-\! C_b ,D\!+\!\sum_{j=1}^K C_j \!-\! C_n;1-C_b,\{1+C_{j}\};\{z_j\} )  
\nonumber \\ &&
+ 
\prod_{j_a=1}^K
   (z_{j_a})^{C_{j_a}} \Gamma\left( -C_{j_a} \right) 
                       \Gamma\left(A \!+\! \sum_{j=1}^K C_{j} \right) 
                       \Gamma\left(D \!+\! \sum_{j=1}^K C_{j} \right)
\nonumber \\ && \hspace{5mm}
            \times
          F_C^{(K)}(A\!+\!\sum_{j=1}^K C_j,D\!+\!\sum_{j=1}^K C_j;\{1+C_{j}\};\{z_j\}) \;. 
\label{Psi<->FC}
\end{eqnarray}
\section{What happens if the monodromy is reducible?}
\label{dimension}
Here, we present a short algebraic deviation of the condition in
Eq.~(\ref{exceptional}) for the system of PDEs related to the Mellin--Barnes integral in Eq.~(\ref{Phi:MB}) to be irreducible and count the dimension of the invariant subspace of the differential contiguous 
operators defined by Eqs.~(\ref{Phi:A:contiguous}) and (\ref{Phi:B:contiguous}).
We follow an idea presented in Ref.~\cite{monodromy}
(see also Ref.~\cite{sadykov}).

Let $\Phi$ be the set of solutions for the system of linear differential operators $L_j^{\Phi}$ 
defined by Eq.~(\ref{sunrise:PDE}) as
$$
L_j^{\Phi}(A,B,\vec{C}): 
\left[ 
\left( \theta_j \!-\! C_j \right) \theta_j  
+
z_j 
\left(\sum_{j=1}^K \theta_j \!+\! A\right)
\left(\sum_{j=1}^K \theta_j \!+\! (1-B)\right)
\right] \;, \quad j = 1, \cdots, K. 
$$
Let $S(A,B,\vec{C})$ denote the local solution space of the operators $L_j^{\Phi}(A,B,\vec{C})$ about some point $z_0$. 
The contiguous differential operators $B^{+}_{A,B,C_j}$ defined by Eq.~(\ref{Phi:contiguous})
map the solution space $S(A,B,\vec{C})$ to the solution 
space  $S(A \pm I_1, B \pm I_2,\vec{C} \pm \vec{I})$, where $\{I_a\}$ are a set of integers.
If the monodromy is reducible, then there is a monodromy-invariant subspace (invariant under the action of the monodromy) 
in the space of solutions.
In this case, the contiguous differential operators $B^{+}_{A,B,C_j}$ have a nontrivial kernel, 
and it is necessary to evaluate their dimension. 

Let us consider the equation $B_A^{+} \Phi =0$, where $B_A^{+}$ is defined by Eq.~(\ref{Phi:A:contiguous}).
Then, the system of equations defined by Eq.~(\ref{sunrise:PDE}) reduces to 
\begin{equation}
L_j^{\Phi}(A,B,\vec{C}) \Phi \equiv 
\left( \theta_j \!-\! C_j \right) \theta_j  \Phi \equiv 0 \;, \quad j = 1, \cdots, K. 
\label{part1}
\end{equation}
If all $C_j \notin \mathbb{Z}$, which is true for the considered Feynman
diagrams, then  
the solution $\Phi_0$ of Eq.~(\ref{part1}) has the following form: 
\begin{eqnarray}
\Phi_0 = c_0 
     + \sum_{i=1}^K c_i z_i^{C_i} 
     + \sum_{\substack{i,j=1 \\ i < j}}^K c_{i,j} z_i^{C_i} z_j^{C_j}
     + \cdots 
     + \mbox{Const.} \times \prod_{i=1}^K z_i^{C_i} \;.
\end{eqnarray}
Applying the operator $B_A^{+}$ to the function $\Phi_0$, we get
\begin{eqnarray}
B_A^{+}\Phi_0 \equiv 0 
& = & 
     A c_0 
     + \sum_{i=1}^K c_i (A + C_i) z_i^{C_i} 
     + \sum_{\substack{i,j=1 \\ i < j}}^K c_{i,j} (A+C_i+C_j) z_i^{C_i} z_j^{C_j}
     + \cdots 
\nonumber \\ && \hspace{-15mm}
     + \left( \prod_{i=1}^K z_i^{C_i}  \right) \times 
        \sum_{a=1}^K  \frac{\tilde{c}_a}{z_a^{C_a}} \left(A \!+\! \sum_{j=1}^K C_j \!-\! C_a \right)
     + \mbox{Const.} \times \prod_{i=1}^K z_i^{C_i} \left( A \!+\! \sum_{j=1}^K C_j \right) \;,
\label{aux:AB}
\end{eqnarray}
where $c_i, c_{i,j}, \ldots, \tilde{c}_i$ are some constants. 
As follows from Eq.~(\ref{aux:AB}),
$B_A^{+} \Phi =0$ if and only if $A+\sum_{S} C_p =0$, where $S$ are any subsets of $\{C_j\}$.
In this way, under the conditions that 
\begin{itemize}
\item  
$A \notin \mathbb{Z},$ 
\item
$C_a \notin \mathbb{Z} \; ,    \quad \forall  a = 1, \cdots, K$,
\item
$A+\sum_{j=1}^K C_j - C_a = 0 \;,  \quad \forall a = 1, \cdots, K$, 
\end{itemize}
there is an invariant subspace of dimension $K$ for the operator $B_A^{+}$.

A similar consideration can also be made for the operator $B_B^{+}$ defined by Eq.~(\ref{Phi:B:contiguous}).
In particular, it is easy to show that, under the conditions that 
\begin{itemize}
\item  
$B \notin \mathbb{Z} \;$, 
\item
$C_a \notin \mathbb{Z} \; , \quad  \forall  a = 1, \cdots, K,$
\item
$B-\sum_{j=1}^K C_j = 0 \;, $
\end{itemize}
there is a one-dimensional invariant subspace.
Collecting the previous results, we get the following lemma: \\
{\it Lemma $\Phi$:}\\
Under the conditions that 
\begin{itemize}
\item
$A$ and $B \notin  \mathbb{Z} \;,$ 
\item
$C_a \notin  \mathbb{Z} \;, \quad  \forall  a = 1, \cdots, K,$
\item
$B-\sum_{j=1}^{K}C_j \in \mathbb{Z} \;,$ \quad 
\item 
$A+\sum_{j=1}^{K}C_j-C_a \in \mathbb{Z} \;, \quad \forall a = 1, \cdots, K \;,$  
\label{LemmaPhi:}
\end{itemize}
there is a $(K+1)$-dimensional invariant subspace (of Puiseux-type solutions)
in the space of solutions of the linear PDEs defined by 
Eq.~(\ref{sunrise:PDE}), and the dimension of the space of nontrivial solutions is equal to 
\begin{eqnarray}
N_\Phi = 2^K-(K+1) \;. 
\label{Phi:rank}
\end{eqnarray}

A similar consideration can also be made for the function $\Psi$ in
Eq.~(\ref{Psi:MB}).

In the application to the Lauricella function $F_C^{(K)}(a,b;\{c_j\};\vec{z})$ of $K$ variables, our analysis is equivalent to the following lemma: \\
{\it Lemma F:}\\
Under the conditions that 
\begin{itemize}
\item
$a$ and $b \notin  \mathbb{Z} \;,$
\item
$c_j \notin  \mathbb{Z} \;, \quad  \forall  ~~j = 1, \cdots, K,$
\item
$b+\sum_{j=1}^{K}c_j \in \mathbb{Z} \;,$ \quad 
\item 
$a+\sum_{j=1}^{K}c_j-c_k \in \mathbb{Z} \;, \quad \forall ~~k = 1, \cdots, K \;,$  
\label{LemmaA:conditions}
\end{itemize}
there is a $(K+1)$-dimensional invariant subspace (of Puiseux-type solutions)
so that the dimension of the space of nontrivial solutions is equal to 
\begin{eqnarray}
N_F = 2^K-(K+1) \;, 
\label{FC:rank}
\end{eqnarray}
and the differential reduction applied to the hypergeometric function 
$F_C^{(K)}(a,b;\{c_j\};\vec{z})$ of $K$ variables has the following form: 
\begin{eqnarray}
F_C^{(K)}(\vec{I}+\{a,b,c_j\};\vec{z}) 
= 
\sum_{j=0}^{K-2} \vec{Q}_j \vec{\theta}^j F_C^{(K)}(a,b;\{c_j\};\vec{z}) 
+ \sum_{j=1}^{K+1} P_j(\vec{z})\;,
\label{red:FC}
\end{eqnarray}
where $\vec{I}$ is a set of integers, $\vec{Q}_j$ are some rational functions,
$P_j(\vec{z})$ denote the Puiseux-type solutions, 
and $\vec{\theta}^p$ means $\theta_{\substack{i_1,i_2,\cdots, i_p \\ i_1 < i_2 < \cdots < i_p}}$. 

The number of symmetric derivatives $\theta_{i_1,i_2,\cdots,i_m}$, 
where $i_1 < i_2 < \cdots < i_m$, is equal to $\frac{K!}{m!(K-m)!}$, 
and the number of terms entering the reduction procedure of Eq.~(\ref{red:FC}) 
is equal to 
$\sum_{j=0}^{K-2} \frac{K!}{j!(K-j)!} = 2^K-(K+1)$, which coincides with Eq.~(\ref{FC:rank}). 
Consequently, the highest differential operators, namely, one operator of order
$K$, $\theta_{i_1,i_2,\cdots,i_K}$,
and $K$ operators of order $K-1$, $\theta_{i_1,i_2,\cdots,i_{K-1}}$,
are expressible in terms of low-dimensional differential operators. 

\noindent
{\it Remark A}: As follows from Eq.~(\ref{aux:AB}), 
for a special set of parameters, the $K+1$ Puiseux-type solutions have the following form: 
$K$ solutions are of the type 
$1/z_a^{C_a} \times \left[ \prod_{j=1}^{K} z_j^{C_j} \right]$, 
and one solution is $\left[ \prod_{j=1}^{K} z_j^{C_j} \right]$.

\noindent
{\it Remark B}: 
For sunrise and bubble diagrams, 
each term of the hypergeometric representation in Eqs.~(\ref{Phi<->FC}) and (\ref{Psi<->FC})
satisfies the conditions of Lemma~F, so that the dimension of the space of nontrivial solutions of each term 
is defined by Eq.~(\ref{FC:rank}) and the differential reduction of each term 
is described by Eq.~(\ref{red:FC}).
%
%

\section{$L$-loop sunrise diagram}
\label{sunrise:L}
The Mellin--Barnes representation for the Feynman diagram defined by Eq.~(\ref{sunset}) 
follows from the algorithm presented in Ref.~\cite{BD} (see also Ref.~\cite{sunrise:2loop})
and has the following form: 
\begin{eqnarray}
&& 
J^{(L)}(M_1^2, \cdots, M_{L+1}^2;  \alpha_1, \cdots, \alpha_{L+1}; p^2)
= 
(p^2)^{\tfrac{n}{2}L-\alpha}
[i^{1-n} \pi^{n/2}]^{L}
\nonumber \\ && 
\times
\int 
\Biggl\{ \prod_{j=1}^{L+1}
dt_j 
\frac{\Gamma(-t_j) \Gamma\left(\frac{n}{2}\!-\!\alpha_j\!-\!t_j \right)}
     {\Gamma(\alpha_j)}
\left( 
- \frac{M_j^2}{p^2}
\right)^{t_j}
\Biggr\}
\frac{
      \Gamma\left( \alpha \!-\! \frac{n}{2}L \!+\! t \right)
     }
     {\Gamma\left( 
      \frac{n}{2}(L+1) \!-\! \alpha \!-\! t \right)
     }
\;, 
\label{mellin-barnes:sunset}
\end{eqnarray}
where 
$$
\alpha  =  \sum_{j=1}^{L+1} \alpha_j \;, 
\quad 
t = \sum_{j=1}^{L+1} t_j \;,
$$
$\alpha_j,L$ are positive integers, $M_j^2$ and $p^2$ 
are some (in general, complex) parameters, and $d$ 
is a parameter (in general, non-integer) of dimensional regularization \cite{dimreg}.

Let us introduce the variables $z_j = - \frac{M_j^2}{p^2}$ ($j=1,2,\cdots,L+1$)
and define the functions $\Phi_J$ as
\begin{equation}
\Phi_J = 
\prod_{k=1}^{L+1} \frac{\Gamma(\alpha_k)}
                     {[i^{1-n} \pi^{n/2}]^L (p^2)^{\frac{n}{2}L-\alpha}}
\times
J^{(L)}(M_1^2, \cdots, M_{L+1}^2;  \alpha_1, \cdots, \alpha_{L+1}; p^2) \;.
\label{Phi_J}
\end{equation}
After this redefinition, the results of Sections~\ref{MB} and \ref{dimension}
are directly applicable to the analysis of the sunrise diagram.
In particular, in the application to the $L$-loop sunrise diagram, we have
\begin{eqnarray}
A = \alpha - \frac{n}{2} L \;, 
\quad 
B = \frac{n}{2} (L+1) - \alpha \;, 
\quad 
C_j = \frac{n}{2} - \alpha_j \;,   \quad j = 1,.\cdots, L+1 \;.
\label{ABC}
\end{eqnarray}
As follows from Eq.~(\ref{aux:AB}), 
to find the dimension of the invariant subspace, it is necessary to find all solutions of the following system of 
algebraic equations:
\begin{subequations}
\begin{eqnarray}
\frac{n}{2}L - \sum_{S_1} \frac{n}{2}  = 0  \quad \left(\mbox{\rm mod} ~ \mathbb{Z} \right) \;,
\label{a}                        
\\ 
\frac{n}{2}(L+1) - \sum_{S_2} \frac{n}{2} = 0 \quad \left(\mbox{\rm mod} ~ \mathbb{Z} \right)\;,
\label{b}
\end{eqnarray}
\label{reducible}
\end{subequations}
\noindent
where $S_1$ and $S_2$ are any subsets of $1,\ldots, L+1$, 
$\mathbb{Z} = \{\ldots, -1, 0, 1, \ldots \}$, and $n$ is non-integer.
The subset $S_1$ in Eq.~(\ref{a}) can be constructed in
$L+1$ different ways (it includes all possible combinations of $L$ out of the
$L+1$ massive lines), 
and there is only one solution of Eq.~(\ref{b}) (the subset $S_2$ includes all the lines). 
In this way, among the $2^{L+1}$ solutions of the system of PDEs related to the sunrise diagram, 
there are $L+1+1$ Puiseux-type solutions.\footnote{As was pointed out in Ref.~\cite{KK2012}, 
a Puiseux-type solution corresponds 
to a product of one-loop bubble diagrams.} Then, the following theorem is valid:  \\
{\bf 
The number $N_J$ of irreducible master integrals 
of the $L$-loop sunrise diagram with generic values of masses and momenta is equal to}
\begin{equation}
\boxed{N_J = 2^{L+1}-L-2} 
\;.
\label{sunset:number}
\end{equation} 
For example, for $L=1,2,3,4,5,6$, we have $N_J = 1,4,11,26,57,120$,
respectively.

The result of the differential reduction can be written in a more familiar form 
via propagators with dots:
the term without derivatives on the r.h.s.\ of Eq.~(\ref{reduction})
corresponds to the diagram itself, whereas terms with derivative(s) $\theta_i \Phi$
correspond to diagrams with dot(s). In the application to the sunrise diagram, we have 
\begin{eqnarray}
J^{(L)}(\{M_i^2\}; \vec{I}+\vec{\alpha}; p^2)
= \sum_{j=0}^{L-1} \vec{Q}_j \vec{\partial}^j 
   J^{(L)}(\{M_i^2\}; \vec{\alpha}; p^2)
+ \sum_{j=1}^{L+1} P_j(\vec{M})\;,
\label{sunrise:reduction}
\end{eqnarray}
where we have introduced the short-hand notation 
$$
J^{(L)}(\{M_i^2\}; \vec{\alpha}; p^2)
\equiv
J^{(L)}(M_1^2, \cdots, M_{L+1}^2;  \alpha_1, \cdots, \alpha_{L+1}; p^2), 
$$
$\vec{Q}_j$ are some rational functions, 
$P_j(\vec{z})$ denote the Puiseux-type solutions, 
and $\vec{\partial}^j$ means symmetric derivatives with respect to the masses
$M^2_j$, i.e.\
\begin{equation}
\vec{\partial}^j \equiv 
\left( 
M^2_{i_1} \frac{\partial}{\partial M^2_{i_1} } 
\right) 
\left( 
M^2_{i_2} \frac{\partial}{\partial M^2_{i_2} }
\right) 
\cdots
\left( 
M^2_{i_j} \frac{\partial}{\partial M^2_{i_j} } 
\right) \;, 
\label{symmetric}
\end{equation} 
and 
$i_1 < i_2 < \cdots < i_j$. 
Indeed, the number of symmetric derivatives $\vec{\partial}^j$ in the considered case is equal to 
$
\frac{(L+1)!}{j!(L+1-j)!} \;
$,
so that 
\begin{equation}
2^{L+1}-(L+2) = \sum_{j=0}^{L-1} \frac{(L+1)!}{j!(L+1-j)!} \;.
\end{equation}
In other words, 
(i) an $L$-loop sunrise diagram  
with two or more derivatives on one of its lines 
is reducible to a linear combination of sunrise diagrams with no more than one
derivative on any of its lines, 
and 
(ii) the number of lines with one dot is less or equal to $L-1$.\footnote{%
Let us recall that there are $L+1$ massive lines in this case.}
The latter statement can also be understood as follows: instead of excluding 
higher-order derivatives, the function (one element) and its first derivatives ($L+1$ elements)
can be excluded in favor of higher derivatives, and the reduction procedure has the following form: 
\begin{subequations}
\begin{eqnarray}
J^{(L)}(\{M_i\},\vec{1};p^2)  
& = & \sum_{j=2}^{L+1} \vec{\tilde{Q}}_j \vec{\partial}^j J^{(L)}(\{M_i\},\{1\};p^2)  
+ \sum_{j=1}^{L+1} \tilde{P}_j(\vec{M})\;,
\\
M_a^2 \frac{\partial}{\partial M_a^2} J^{(L)}(\{M_i\},\vec{1};p^2)  
& = & \sum_{j=2}^{L+1} \vec{\hat{Q}}_j \vec{\partial}^j J^{(L)}(\{M_i\},\{1\};p^2)  
+ \sum_{j=1}^{L+1} \hat{P}_j(\vec{M})\;, \quad \forall a = 1, \cdots, L+1, 
\nonumber \\ && 
\\ 
J^{(L)}(\{M_i\},\vec{I}+\{\alpha\};p^2)  
& = & \sum_{j=2}^{L+1} \vec{Q}_j \vec{\partial}^j J^{(L)}(\{M_i\},\{1\};p^2)  
+ \sum_{j=1}^{L+1} P_j(\vec{M})\;,
\\
\end{eqnarray}
\label{sunrise:reduction:symmetric}
\end{subequations}
\noindent
where $\vec{Q}_j, \vec{\tilde{Q}}, \vec{\hat{Q}}$ are some rational functions,
$P_j(\vec{z}), \tilde{P}_j(\vec{M}), \hat{P}_j(\vec{M})$ denote the Puiseux-type
solutions, and $\vec{\partial}^j$ is defined by Eq.~(\ref{symmetric}).
As an illustration of these relations, let us consider some special cases.
\begin{itemize}
\item
At the two-loop level $(L=2)$, only first derivatives with respect to masses 
enter the reduction procedure, in agreement with
Refs.~\cite{tarasov,DS99,ita}.
\item At the three-loop level $(L=3)$, 
according to Eq.~(\ref{sunrise:reduction}),  
the second symmetric derivatives with respect to masses 
$\frac{\partial^2}{\partial M^2_i \partial M_j^2} J^{(3)}({M_i^2};\vec{1};p^2)$ 
where $i<j$ and $i,j=1,2,3,4$, are generated ($1,4,6$ terms)
or, according to Eq.~(\ref{sunrise:reduction:symmetric}), the basis can
be 
constructed from the second, third, and fourth symmetric derivatives ($6,4,1$ terms), 
$\frac{\partial^2}{\partial M^2_i \partial M_j^2} J^{(3)}({M_i^2};\vec{1};p^2)$, 
$\frac{\partial^3}{\partial M^2_i \partial M_j^2 \partial M_k^2} J^{(3)}({M_i^2};\vec{1};p^2)$, 
and 
$\frac{\partial^4}{\partial M^2_1 \partial M_2^2 \partial M^2_3 \partial M_4^2} J^{(3)}({M_i^2};\vec{1};p^2)$, 
where $i<j<k$ and $i,j,k=1,2,3,4$.
\item
At the four-loop level $(L=4)$,
according to Eq.~(\ref{sunrise:reduction}),  
the third symmetric derivatives with respect to masses 
$\frac{\partial^3}{\partial M^2_i \partial M_j^2 \partial M_k^2} J^{(4)}({M_i^2};\vec{1};p^2)$, 
where $i<j<k$ and $i,j,k=1,2,3,4,5$, are generated ($1,5,10,10$ terms)
or, according to Eq.~(\ref{sunrise:reduction:symmetric}), the basis includes
the second, third, fourth, and fifth symmetric derivatives 
($10,10,5,1$ terms),
$\frac{\partial^2}{\partial M^2_i \partial M_j^2} J^{(4)}({M_i^2};\vec{1};p^2)$,
$\frac{\partial^3}{\partial M^2_i \partial M_j^2 \partial M_k^2} J^{(4)}({M_i^2};\vec{1};p^2)$,
$\frac{\partial^4}{\partial M^2_i \partial M^2_j \partial M^2_k \partial M_l^2} J^{(4)}({M_i^2};\vec{1};p^2)$,
and 
$\frac{\partial^5}{\partial M^2_1 \partial M_2^2 \partial M^2_3 \partial M_4^2 \partial M_5^2} 
     J^{(4)}({M_i^2};\vec{1};p^2)$,
where $i<j<k<l$ and $i,j,k,l=1,2,3,4,5$.
\end{itemize}

The case of integer $n$ requires an extra analysis \cite{sunrise:integer}. \\


\subsection{$L$-loop sunrise diagram with $R$ massive lines}
Let us consider the $L$-loop sunrise diagram in which only $R$ lines $(R \leq L)$ have different masses,
\begin{eqnarray}
&& \hspace{-5mm}
J_{R}(\{M_{i}^2\}; \{\alpha_j\}, \{\beta_k \}; p^2)
= 
\int \prod_{j=1}^R \frac{d^n (k_1 \cdots k_L)}
{
[k_j^2\!-\!M_k^2]^{\alpha_j} 
[k_{R+1}^2]^{\beta_1} 
\cdots
[(p\!-\!k_1\!-\cdots\!-\!k_L)^2]^{\beta_{L+1-R}}   
} 
\;. 
\nonumber \\ && 
\label{sunset:R}
\end{eqnarray}
The Mellin--Barnes representation of this diagram is 
\begin{eqnarray}
&& 
J_{R}(\{M_{R}^2\}; \{\alpha_j\}, \{\beta_k \}; p^2)
= 
(p^2)^{\tfrac{n}{2}L-\alpha-\beta}
[i^{1-n} \pi^{n/2}]^{L}
\nonumber \\ && 
\times
\Biggl\{ \prod_{k=1}^{L+1-R}
\frac{\Gamma\left(\frac{n}{2}\!-\!\beta_k\! \right)}
     {\Gamma(\beta_k)}
\Biggr\}
\frac{\Gamma\left(\beta\!-\!\frac{n}{2}(L\!-\!R)\right)}
     {\Gamma\left(\frac{n}{2}(L\!-\!R\!+\!1)\!-\!\beta\right)}
\nonumber \\ && 
\times
\int 
\Biggl\{ \prod_{j=1}^{R}
dt_j 
\frac{\Gamma(-t_j) \Gamma\left(\frac{n}{2}\!-\!\alpha_j\!-\!t_j \right)}
     {\Gamma(\alpha_j)}
\left( 
- \frac{M_j^2}{p^2}
\right)^{t_j}
\Biggr\}
\frac{
      \Gamma\left( \alpha \!+\! \beta \!-\! \frac{n}{2}L \!+\! t \right)
     }
     {\Gamma\left( 
      \frac{n}{2}(L+1) \!-\! \alpha \!-\! \beta \!-\! t \right)
     }
\;, 
\label{mellin-barnes:sunset:R}
\end{eqnarray}
where 
$$
\alpha  =  \sum_{j=1}^{R} \alpha_j \;, 
\quad 
\beta  =  \sum_{j=1}^{L-R+1} \beta_j \;, 
\quad 
t = \sum_{j=1}^{R} t_j \;.
$$
In this case, we have:
$$
A = \alpha \!+\! \beta \!-\! \frac{n}{2} L \;, 
\quad 
B = \frac{n}{2}(L+1) \!-\! \alpha \!-\! \beta \;, 
\quad 
C_j = \frac{n}{2} -\alpha_j \;, \quad j=1, \cdots, R.    
$$
To find the dimension of the invariant subspace, we use the algorithm described in Section \ref{dimension}. 
In this case, it is necessary to find all solutions (all subsets) of the following system of algebraic equations:
\begin{subequations}
\begin{eqnarray}
\frac{n}{2}L - \sum_{S_1} \frac{n}{2}  = 0   \quad \left(\mbox{\rm mod} ~~ \mathbb{Z} \right) \;,
\label{I}
\\ 
\frac{n}{2}(L+1) - \sum_{S_2} \frac{n}{2} = 0 \quad \left(\mbox{\rm mod} ~~ \mathbb{Z} \right) \;,
\label{II}
\end{eqnarray}
\end{subequations}
where $S_1$ and $S_2$ are any subsets of $1,\ldots, R$, and $n$ is non-integer.
There is only one solution for the subset $S_1$ (if $R=L$), and there is no solution for Eq.~(\ref{II}).
In this way, among the $2^{R}$ solutions, there is only one Puiseux-type
solution (if $R=L$).
Then we have the following theorem: \\
{\bf  
The number $N_{J,R}$ of irreducible master integrals 
of the $L$-loop sunrise diagram with $R$ massive lines $(R \leq L)$ is equal to
} 
\begin{equation}
\boxed{N_{L,R} = 2^{R}-\delta_{0,L-R}} \;.
\label{sunset:number:R}
\end{equation} 
As follows from this relation,
the sunrise diagrams with $R$ massive and two or more massless lines
are irreducible, and their holonomic ranks near $\vec{z} =0$ 
coincide with the holonomic ranks of the hypergeometric functions $F_C^{(R)}$ 
with irreducible monodromies.

There is a one-dimensional invariant subspace (Puiseux-type solution)
if the sunrise diagram has one massless line.\footnote{At the two-loop level,
this was shown 
explicitly by Tarasov in Ref.~\cite{tarasov} (see also the discussion in
Ref.~\cite{DS99}).}
If at least one of the values of $\beta_j$ is non-integer, which happens if a
massless line is dressed by other massless lines, then the 
number of irreducible master integrals is equal to $2^{R}$. \\

\noindent
{\it Example:} 
$$
N_{1,1}=1 \; ;
\quad 
N_{2,2}=3 \;, 
\quad   
N_{2,1}=2  \;; 
\quad 
N_{3,3}=7  \;, 
\quad   
N_{3,2}=4  \;,  
\quad   
N_{3,1}=2  \;. 
$$

\section{$L$-loop bubble diagram}

\begin{figure}[t]
\centering 
\includegraphics[width=60mm,height=80mm]{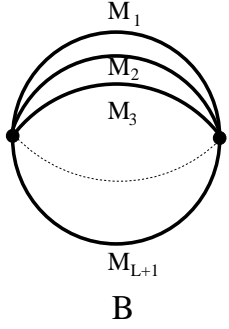}
\caption{\label{bubblefig}
$L$-loop bubble diagram.
}
\end{figure}

In a similar manner, let us consider the $L$-loop bubble diagram 
(see Fig.~\ref{bubblefig}) defined as
\begin{eqnarray}
&& \hspace{-5mm}
B(M_1^2, \cdots, M_{L+1}^2;  \alpha_1, \cdots, \alpha_{L+1})
= 
\int \frac{d^n (k_1 \cdots k_L)}
{
[k_1^2\!-\!M_1^2]^{\alpha_1} 
\cdots
[k_L^2\!-\!M_L^2]^{\alpha_L} 
[(\!k_1\!-\cdots\!-\!k_L)^2\!-\!M_{L+1}^2]^{\alpha_{L+1}}  
} \;,
\nonumber \\ && 
\label{bubble}
\end{eqnarray}
where $\alpha_j$ are integers. 
Its Mellin--Barnes representation is 
\begin{eqnarray}
&& 
B(M_1^2, \cdots, M_{L+1}^2;\alpha_1, \cdots, \alpha_{L+1})
= 
(-M^2_{L+1})^{\tfrac{n}{2}L-\alpha}
\times 
\frac{[i^{1-n} \pi^{n/2}]^{L}}
     {\Gamma\left(\tfrac{n}{2}\right) \Gamma(\alpha_{L+1})}
\nonumber \\ && \hspace{-7mm}
\times
\int 
\Biggl\{ \prod_{j=1}^{L}
dt_j 
\frac{\Gamma(-t_j) \Gamma\left(\frac{n}{2}\!-\!\alpha_j\!-\!t_j \right)}
     {\Gamma(\alpha_j)}
\left( 
\frac{M_j^2}{M^2_{L+1}}
\right)^{t_j}
\Biggr\}
      \Gamma\left( \alpha \!+\! \alpha_{L+1} \!-\! \frac{n}{2}L \!+\! t \right)
      \Gamma\left( \alpha \!-\! \frac{n}{2}(L\!-\!1) \!+\! t   \right)
\;, 
\nonumber \\ && 
\label{mellin-barnes:bubble}
\end{eqnarray}
where 
$$
\alpha  =  \sum_{j=1}^{L} \alpha_j \;, 
\quad 
t = \sum_{j=1}^{L} t_j \;.
$$
Let us introduce the variables $z_j = \frac{M_j^2}{M_{L+1}}$, $j=1,\cdots,K,$
and define the functions $\Psi_B$ as
$$
\Psi_B = 
\prod_{k=1}^{L+1} \Gamma(\alpha_k) \times 
                \frac{\Gamma\left(\frac{n}{2}\right)}
                     {[i^{1-n} \pi^{n/2}]^L (-M_{L+1}^2)^{\frac{n}{2}L-\alpha}}
\times
B(M_1^2, \cdots, M_{L+1}^2; \alpha_1, \cdots, \alpha_{L+1}) \;.
$$
In this case, we have
$$
A = \alpha \!+\! \alpha_{L+1} \!-\! \frac{n}{2} L \;, 
\quad 
D = \alpha \!-\! \frac{n}{2}(L-1) \;, 
\quad 
C_j = \frac{n}{2} -\alpha_j \;,   \quad j=1, \cdots, L \;.
$$
As follows from Eq.~(\ref{aux:AB}), the dimension of the invariant subspace
it defined by the number of solutions of the following system of algebraic equations:
\begin{subequations}
\begin{eqnarray}
\frac{n}{2}L - \sum_{S_1} \frac{n}{2}  = 0   \quad \left(\mbox{\rm mod} ~~~ \mathbb{Z} \right) \;,
\label{tad1}
\\ 
\frac{n}{2}(L-1) - \sum_{S_2} \frac{n}{2} = 0 \quad \left(\mbox{\rm mod} ~~~ \mathbb{Z} \right) \;,
\label{tad2}
\end{eqnarray}
\end{subequations}
where $S_1$ and $S_2$ are any subsets of $1,\ldots, L$, and $n$ is non-integer.
There is only one solution for the subset $S_1$, and there are $L$ solutions for Eq.~(\ref{tad2}).
In this case, there are $L+1$ Puiseux-type solutions, 
and the following theorem is valid:\\
{\bf 
The number $N_B$ of irreducible master integrals 
for the $L$-loop bubble diagram with generic values of masses is equal to} 
\begin{equation}
\boxed{N_B = 2^{L}-L-1} \;.
\label{bubble:number}
\end{equation} 
For example, for $L=1,2,3,4,5,6$, there are $N_B=0,1,4,11,26,57$ nontrivial master integrals.
Let us explain the zero result for $L=1$. The one-loop tadpole $A_0$ 
is treated as a constant in the framework of the differential algebra (see the discussion in Refs.~\cite{KK2012,counting}).

{\it Remark 1}: The result in Eq.~(\ref{bubble:number}) can be derived from
Eq.~(\ref{sunset:number}) by considering bubble diagrams as sunrise diagrams with external momenta put 
equal to zero, which effectively reduces the number of independent variables by one.

{\it Remark 2}: In contrast to the sunrise diagram, 
the reduction for the bubble diagrams with $L \geq 3$ 
does not have a completely symmetric structure with respect to mass derivatives.
At the three-loop level, this statement was confirmed in Ref.~\cite{spmartin}. 
In the framework of integration-by-parts (IBP) relations \cite{ibp}, there is a
so-called $\{{\tt dim}\}$ relation 
(see the discussion in Ref.~\cite{avdeev}) connecting the diagram in
Eq.~(\ref{bubble}) without derivatives
to a linear combination of diagrams with a first derivative,  
\begin{equation}
\left\{ 
\sum_{j=1}^{L+1} M_j^2 \frac{\partial}{\partial M_j^2} 
- 
\left(\frac{n}{2}L - \sum_{j=1}^{L+1} \alpha_j \right) 
\right\} 
B(M_1^2, \cdots, M_{L+1}^2;\alpha_1, \cdots, \alpha_{L+1})
= 0 \;,
\label{dim}
\end{equation}
so that 
\begin{equation}
B(\{M_{L+1}^2\}; \vec{1})
= \frac{1}{\left[ \frac{n}{2}L-(L+1) \right]} 
\left( \sum_{j=1}^{L+1} B(\{M_{i}^2\}; \vec{1}+\vec{e_j}) \right)  \;,
\label{B-unit}
\end{equation}
where $\vec{e_j}$ is the unit vector with unity in the $j$-th place and we have
introduced the short-hand notation
$
B(\{M_{L+1}^2\}; \vec{1})
\equiv
B(M_1^2, \cdots, M_{L+1}^2;1, \cdots, 1) \;.
$

{\it Remark 3}: 
Let us find out which diagrams form the symmetric set of master integrals 
for the bubble diagrams with $L \geq 4$.
It is easy to see that the lists of diagrams are different for even and odd
numbers of loops.
If $L$ is even, then the symmetric set of master integrals is defined by the following expression:
\begin{eqnarray}
\sum_{j=0}^{L-2} \frac{(L+1)!}{j!(L+2-j)!} \frac{(1+(-1)^j)}{2}\vec{\partial}^j 
B(M_1^2, \cdots, M_{L+1}^2;1,\cdots,1) \;,
\label{bubble:even}
\end{eqnarray}
where $\vec{\partial}^j$ is defined by Eq.~(\ref{symmetric}).
For odd values of $L$, a similar expression is valid, namely 
\begin{eqnarray}
\sum_{j=0}^{L-2} \frac{(L+1)!}{j!(L+2-j)!} \frac{(1-(-1)^j)}{2}
\vec{\partial}^j B(M_1^2, \cdots, M_{L+1}^2; 1, \cdots,1) \;.
\label{bubble:odd}
\end{eqnarray}
As an illustration of these relations, let us consider some special cases.
\begin{itemize}
\item
At the four-loop level $(L=4)$, there are 11 irreducible master integrals,
and the symmetric set of master integrals includes the diagram 
$B(\{M_i^2\};\{1\})$ (1 term)
and its second symmetric derivatives with respect to masses, 
$\frac{\partial^2}{\partial M^2_i \partial M_j^2} B(\{M_i^2\};\{1\})$,
where $i<j$ and $i,j=1,\cdots,5$ ($10$ terms).
\item
At the five-loop level $(L=5)$, 
there are 26 irreducible master integrals,
and the symmetric set of master integrals include 
their first derivatives with respect to masses, 
$\frac{\partial}{\partial M^2_i} B(\{M_i^2\};\{1\})$, 
where $j=1,\cdots,6$ ($6$ terms), 
and their third symmetric derivatives with respect to masses,
$\frac{\partial^3}{\partial M^2_i \partial M_j^2 \partial M_k^2} B(\{M_i^2\};\{1\})$, 
where $i<j<k$ and $i,j,k=1,\cdots,6$ (20 terms).
\item
At the six-loop level $(L=6)$,
there are 57 irreducible master-integrals, 
and the symmetric set of master integrals 
include the diagram $B(\{M_i^2\};\{1\})$ ($1$ term), 
its second and fourth symmetric derivatives with respect to masses, 
$\frac{\partial^2}{\partial M^2_i \partial M_j^2} B(\{M_i^2\};\{1\})$
and 
$\frac{\partial^4}{\partial M^2_i \partial M_j^2 \partial M_k^2 \partial M_r^2 } B(\{M_i^2\};\{1\})$, 
where $i<j<k<r$ and $i,j,k,r=1,\cdots,7$ ($21$ and $35$ terms, respectively).
\end{itemize}
\subsection{$L$-loop bubble diagram with $R$ massive lines}
Let us consider the $L$-loop bubble diagram for the case where only $R$ lines $(R \leq L)$ have different masses,
\begin{eqnarray}
&& \hspace{-5mm}
B_{R}(M_1^2, \cdots, M_{R}^2;  \{ \alpha_j \}, \{ \beta_k \})
= 
\int \frac{d^n (k_1 \cdots k_L)}
{
[k_1^2\!-\!M_1^2]^{\alpha_1} 
\cdots
[k_R^2\!-\!M_R^2]^{\alpha_R} 
[k_{R+1}^2]^{\beta_1} 
\cdots
[(\!k_1\!-\cdots\!-\!k_L)^2]^{\beta_{L+1-R}}   
} 
\;. 
\nonumber \\ && 
\label{bubble:R}
\end{eqnarray}
The Mellin--Barnes representation of this diagram is 
\begin{eqnarray}
&& 
B_{R}(M_1^2, \cdots, M_{R}^2;  \{ \alpha_j \}, \{ \beta_k \})
= 
(-M_R^2)^{\tfrac{n}{2}L-\alpha-\beta}
\frac{[i^{1-n} \pi^{n/2}]^{L}}
     {\Gamma\left(\tfrac{n}{2}\right) \Gamma(\alpha_{L+1})}
\times 
\Biggl\{ \prod_{k=1}^{L+1-R}
\frac{\Gamma\left(\frac{n}{2}\!-\!\beta_k\! \right)}
     {\Gamma(\beta_k)}
\Biggr\}
\nonumber \\ && 
\times
\frac{\Gamma\left(\beta\!-\!\frac{n}{2}(L\!-\!R)\right)}
     {\Gamma\left(\frac{n}{2}(L\!-\!R\!+\!1)\!-\!\beta\right)}
\times
\int 
\Biggl\{ \prod_{j=1}^{R-1}
dt_j 
\frac{\Gamma(-t_j) \Gamma\left(\frac{n}{2}\!-\!\alpha_j\!-\!t_j \right)}
     {\Gamma(\alpha_j)}
\left( 
\frac{M_j^2}{M_R}
\right)^{t_j}
\Biggr\}
\nonumber \\ && 
\times
      \Gamma\left( \alpha \!+\! \beta \!-\! \frac{n}{2}(L-1) \!+\! t \right)
      \Gamma\left( \alpha \!+\! \beta \!+\! \alpha_{R} \!-\! \frac{n}{2}L \!+\! t \right)
\;, 
\nonumber \\ && 
\label{mellin-barnes:bubble:R}
\end{eqnarray}
where 
$$
\alpha  =  \sum_{j=1}^{R-1} \alpha_j \;, 
\quad 
\beta  =  \sum_{j=1}^{L-R+1} \beta_j \;, 
\quad 
t = \sum_{j=1}^{R-1} t_j \;.
$$
In terms of the notations of Section~\ref{MB}, we have
$$
A = \alpha \!+\! \beta \!-\! \frac{n}{2} (L-1) \;, 
\quad 
D = \alpha \!+\! \beta \!+\! \alpha_R \!-\! \frac{n}{2}L \;, 
\quad 
C_j = \frac{n}{2} -\alpha_j \;, \quad j=1, \cdots, R-1.   
$$
Applying again the algorithm described in Section~\ref{dimension}, we get
\begin{subequations}
\begin{eqnarray}
\frac{n}{2}(L-1) - \sum_{S_1} \frac{n}{2}  = 0  \quad \left(\mbox{\rm mod} ~~~ \mathbb{Z} \right) \;,
\label{x}
\\ 
\frac{n}{2}L - \sum_{S_2} \frac{n}{2}      = 0  \quad \left(\mbox{\rm mod} ~~~ \mathbb{Z} \right) \;,
\label{y}
\end{eqnarray}
\end{subequations}
where $S_1$ and $S_2$ are any subsets of $1,\ldots, R-1$, and $n$ is
non-integer.
There is only one solution for the subset $S_1$ (if $R=L$), and there is no solution for Eq.~(\ref{y}).
In this way, among the $2^{R-1}$ solutions, there is only one Puiseux-type solution (if $R=L$).
Then we have the following theorem: \\
{\bf The number $B_{L,R}$ of irreducible master integrals 
of the $L$-loop bubble diagram with $R$ massive lines $(R \leq L)$ is equal to} 
\begin{equation}
\boxed{B_{L,R} = 2^{R-1}-\delta_{0,L-R}} \;.
\label{bubble:number:R}
\end{equation} 
{\it Example}: 
$$
B_{2,2}=1 \;, \quad
B_{3,3}=3 \;, \quad  
B_{3,2}=2 \;, \quad  
B_{4,4}=7  \;, \quad   
B_{4,3}=4 \;, \quad    
B_{4,2}=2 \;. 
$$

\section{Independent verification}
\label{independent}

Fortunately, there are a few others ways to cross-check our key results, 
Eqs.~(\ref{sunset:number}), (\ref{sunset:number:R}), (\ref{bubble:number}), and
(\ref{bubble:number:R}).
One is based on the reduction of sunrise and bubble diagrams to some bases by using IBP relations \cite{ibp}.
However, there is no guarantee that the freely available programs will perform
the complete reduction
(see the discussions in Refs.~\cite{KK:2011,KT:2016,Boels:2015}). 

Another way of cross-checking our result is to apply the Lee-Pomenransky
approach \cite{Mint} (see also the discussion in Ref.~\cite{baikov}).
This algorithm is based on counting the critical points of the sum of the
Symanzik polynomials $F+U$ defined by Eq.~(\ref{sunrise:parametric}). As was
pointed out by the authors of Ref.~\cite{Mint},
there are situations where their program 
does not reproduce the correct number of master integrals. Here, we present a
pedagogical example in which
differential reduction allows us to predict and construct an algebraic relation 
between two master integrals \cite{KK:2011},
which is not predictable by the algorithm described in Ref.~\cite{Mint} 
or by the program {\tt LiteRed} \cite{LiteRed}.\footnote{
For the considered example, the diagram ${\tt J011}$, 
the program {\tt Azurite} \cite{Azurite} also produces two irreducible master
integrals.
We thank the authors of Ref.~\cite{Azurite} for confirming and verifying this
result.}
Let us consider the two-loop sunrise diagram with arbitrary kinematics. 
The sum of the corresponding Symanzik polynomials has the following form:
$$
G \equiv 
F + U = z_1 z_2 + z_1 z_3 + z_2 z_3 
+ z_1 z_2 z_3 p^2 
- 
\left( z_1 z_2 + z_1 z_3 + z_2 z_3 \right) 
\left( z_1 M_1^2 + z_2 M_2^2 + z_3 M_3^2 \right)  \;.
$$
In this case, there are eight critical points, defined by the conditions
$G=0$ and $\partial_{z_i} G = 0$, $i =1,2,3$.
Four of them are trivial, $G(\{q_1,q_2,q_3,q_4\}) = 0$,
where $q_1=(1/M_1^2,0,0)$, $q_2=(0,1/M_2^2,0)$, $q_3=(0,0,1/M_3^2)$, and
$q_4 = (0,0,0)$. 
The remaining four points are algebraically independent for a generic set of
masses and momenta, 
so that there are four independent master integrals, in agreement with the result of Ref.~\cite{tarasov} . 
The number and the values of critical points do not 
depend on the values of $\alpha_i$ (power of propagators) and the dimension of space-time $n$, 
and the product of one-loop bubble diagrams does not enter the counting of master integrals. 

Let us consider as another particular case the two-loop on-shell diagram, which 
is denoted as {\tt J011} in Ref.~\cite{KK:2011}, with
$M_3=0$ and all other masses on-mass shell, $M_1^2=M_2^2=p^2=1$. 
In this case, there are six non-degenerate\footnote{%
The  corresponding Hessian matrix is non-singular at these points.}
critical points, 
\begin{eqnarray}
&& 
q_1=(0,0,0), 
\quad 
q_2=(1,0,0), 
\quad 
q_3=(0,1,0),   
\nonumber \\ && 
q_4=\left(-\frac{1}{3}, \frac{2}{3}, -\frac{2}{3} \right),
\quad
q_5=\left(\frac{2}{3}, -\frac{1}{3}, -\frac{2}{3} \right),  
\quad 
q_6=\left(\frac{2}{3}, \frac{2}{3}, -\frac{2}{3} \right),  
\nonumber 
\end{eqnarray}
and 
$$
G(q_1) = G(q_2) = G(q_3) = 0; 
\quad 
G(q_4) = G(q_5) = G(q_6) = -\frac{4}{27}. 
$$
The index\footnote{
The index of a critical point is the dimension of the negative eigenspace
of the corresponding Hessian matrix at this point.} 
of points $q_1=q_2=q_3$ is equal to 2,  
and the index of points $q_4=q_5=q_6$ is equal to 1.  
According to the criteria suggested in Ref.~\cite{Mint}, 
there are two independent critical points, $q_4$ and $q_6$,
and there are two master integrals which are not reducible to products of one-loop 
bubble diagrams.
However, as was shown in Refs.~\cite{KK:2011,additional:2loop}, 
there is only one non-trivial master integral in this case,
the second one only being a product of $\Gamma$ functions.\footnote{%
Roman Lee made the following comment on this: since the algebraic relation between the master integrals 
established in Refs.~\cite{KK:2011,KT:2016} does not follow from IBP relations related to the sunrise diagram 
\cite{additional:2loop}, our results do not contradict the algorithm described in Ref.~\cite{Mint}.}  

The difference between the differential-reduction technique and the counting of
the critical points is related to the treatment of the values of the propagator
powers and the space-time dimension. 
In the framework of the differential-reduction technique, one can consider the
parameters $\alpha_i$ (powers of propagators) as well as the dimension of space-time $n$ as non-integer, and, as a consequence, 
the number of additional differential equations for the Mellin--Barnes
integrals 
should be different in this case.  In the Lee-Pomeransky approach, the dimension of 
space-time and the powers of the propagators do not enter the analysis. 
Nevertheless, if $\alpha_i$ are integer, then the results of our evaluations and
those of the application of {\tt Mint} should be equal. 
Roman Lee kindly agreed to cross-check our results by the help of his packages
\cite{Mint,LiteRed} 
and got full agreement with our expressions through $L=5$ for the sunrise
diagrams and through $L=6$ for the bubble diagrams.

\section{Conclusions}
The method of counting master integrals described in
Refs.~\cite{counting,KK2012} 
was applied to the multivariable case with reducible monodromy. 
In contrast to our previous considerations \cite{diagrams,KK:2011}, 
the case where some additional differential equations are generated simultaneously 
was analysed in the present paper.  
Our technique is based on the methods developed for the analysis of the 
monodromy of GKZ hypergeometric functions \cite{beukers,GKZ:monodromy,monodromy,sadykov}.

For completeness, we recall the three basic steps of our algorithm: 
(i) get a system of linear PDEs
for a given Feynman diagram via its Mellin-Barnes 
representation, which does not exploit IBP relations \cite{ibp};
(ii) 
evaluate the holonomic rank 
(including zero and Puiseux-type solutions) of the system of PDEs
by the help of the prolongation procedure; 
(iii) evaluate the dimension of the invariant subspace of the differential 
contiguous operators. 

%
%
%
%
%
%
%
%

As a demonstration of the validity of this technique applied to Feynman
diagrams, 
the numbers of irreducible master integrals for the $L$-loop sunrise and bubble 
diagrams with generic sets of masses and momenta were evaluated
(see Eqs.~(\ref{sunset:number}), (\ref{sunset:number:R}), (\ref{bubble:number}),
and (\ref{bubble:number:R})).
In the considered cases, the non-integer value of the space-time dimension $n$ serves as a regulator 
which allows us to count the number of independent solutions. 

As a by-product, we discovered the following interesting consequences of our analysis:
\begin{itemize}
\item
As follows from Eq.~(\ref{sunrise:reduction}), 
the bases for $L$-loop sunrise diagrams can be constructed in two equivalent ways.
In the first set, sunrise diagrams with higher symmetric derivatives with respect to masses,  
$\frac{\partial^{J}}{\partial M^2_{i_1}  \partial M^2_{i_2} \cdots \partial M^2_{i_{J}}} 
 J^{(L)}(\vec{M_j^2};\vec{1};p^2)$,
where $J=L,L+1$ and $i_1 < i_2 < \cdots < i_J$, can be excluded (see Eq.~(\ref{sunrise:reduction})).
In the second set, the diagram with unit powers of propagators and its first derivatives with respect to 
masses can be excluded in favor of higher derivatives (see Eq.~(\ref{sunrise:reduction:symmetric})).
In both cases, the numbers of basic elements coincide with Eq.~(\ref{sunset:number}),
and the bases do not include the diagrams with powers of propagator larger than $2$.
\item
As follows from Eq.~(\ref{bubble:number}), the bases for $L$-loop bubble 
diagrams have the following structure: 
for even numbers of loops $L=2,4,6, \cdots$, the basis includes the diagram 
with unit powers of propagators, $B(\{M_i^2\};\{1\})$, and its 
even-order symmetric derivatives with respect to masses (see Eq.~(\ref{bubble:even})).
For odd numbers of loops $L=3,5,7, \cdots$, the basis includes only 
odd-order symmetric derivatives with respect to masses (see Eq.~(\ref{bubble:odd})),
but does not include the original diagram.
\item
As follows from Eq.~(\ref{sunset:number:R}), 
a sunrise diagram with $R$ massive and two or more massless lines, 
or with one ``dressed'' massless line, 
is irreducible, and its holonomic rank near $\vec{z} =0$ 
coincides with the holonomic rank of the hypergeometric function $F_C^{(R)}$ 
with irreducible monodromy.
The latter property is relevant for the analysis of special functions 
generated by the $\ep$ expansions of multiloop sunrise diagrams \cite{borwein,sunrise,watermellon}.
\end{itemize}

\appendix
\section{$L$-loop V-type diagram}
\label{appendixA} 
\begin{figure}[th]
\centering 
\includegraphics[width=60mm,height=80mm]{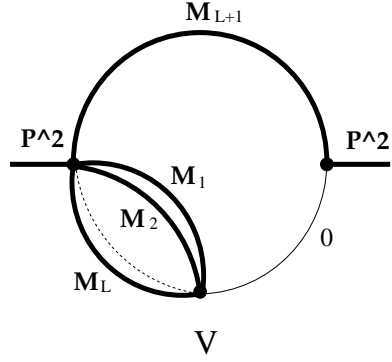}
\caption{\label{Vfig}
$L$-loop V-type diagram.
}
\end{figure}

In recent analyses aiming at finding the full sets of IBP relations for the
complete reduction of Feynman diagram to minimal sets of master integrals
\cite{KK2012,KT:2016,additional:3loop,additional:2loop}, 
diagrams of $V$ type (see Fig.~\ref{Vfig}) have played an important role.
In this section, we prove the following theorem:  \\
\noindent 
{\bf 
  For generic values of masses and external momenta, the $V$-type diagrams
  (see Fig.~\ref{Vfig}) are reducible to sunrise diagrams with the same masses and
momenta. 
}

Let us consider the $L$-loop V-type diagram defined in momentum space as
\begin{eqnarray}
&& \hspace{-5mm}
V^{(L)}(M_1^2, \cdots, M_{L+1}^2,  \alpha_1, \cdots, \alpha_{L+1}; \sigma; p^2)
\nonumber \\ && 
= 
\int \frac{d^n (k_1 \cdots k_L)}
{
[k_1^2\!-\!M_1^2]^{\alpha_1} 
\cdots
[(k_1\!-\cdots\!-\!k_L)^2\!-\!M_{L}^2]^{\alpha_{L}}   
[(p\!-\!k_L)^2\!-\!M_{L+1}^2]^{\alpha_{L+1}}   
[k_L^2]^{\sigma}   
} 
\;. 
\nonumber \\ && 
\label{V}
\end{eqnarray}
The Mellin--Barnes representation of this diagram is 
\begin{eqnarray}
&& 
V^{(L)}(M_1^2, \cdots, M_{L+1}^2,  \alpha_1, \cdots, \alpha_{L+1}; \sigma; p^2)
= 
(p^2)^{\tfrac{n}{2}L-\alpha-\alpha_{L+1}-\beta-\sigma}
[i^{1-n} \pi^{n/2}]^{L}
\nonumber \\ && 
\times
\int 
\Biggl\{ \prod_{j=1}^{L+1}
dt_j  
\frac{\Gamma(-t_j) \Gamma\left(\frac{n}{2}\!-\!\alpha_j\!-\!t_j \right)}
     {\Gamma(\alpha_j)}
\left( 
- \frac{M_j^2}{p^2}
\right)^{t_j}
\Biggr\}
\frac{
      \Gamma\left( \frac{n}{2}L \!-\! \alpha \!-\! \sigma \!-\! t \right)
      \Gamma\left( \alpha \!-\! \frac{n}{2}(L-1) \!+\! t \right)
     }
     {\Gamma\left( \frac{n}{2}L \!-\! \alpha \!-\! t \right)
      \Gamma\left( \alpha \!+\! \sigma \!-\! \frac{n}{2}(L-1) \!+\! t \right)
     } 
\nonumber \\ && 
\times
\frac{
      \Gamma\left( \alpha \!+\! \alpha_{L+1} \!+\! \sigma \!-\! \frac{n}{2}L \!+\! t \!+\! t_{L+1}\right)
     }
     {
      \Gamma\left( \frac{n}{2}(L+1) \!-\! \alpha \!-\! \alpha_{L+1} \!-\! \sigma \!-\! t \!-\! t_{L+1}\right)
     }
\;, 
\label{mellin-barnes:v}
\end{eqnarray}
where 
$$
L \geq 2 \;, 
\quad 
\alpha  =  \sum_{j=1}^{L} \alpha_j \;, 
\quad 
t = \sum_{j=1}^{L} t_j \;.
$$
For $\sigma=0$, the V-type diagram coincides with the $L$-loop sunrise diagram.
The integral in Eq.~(\ref{mellin-barnes:v})
includes the following ratio of $\Gamma$ functions with arguments differing by
integers:
$$
\frac{
      \Gamma\left( \frac{n}{2}L \!-\! \alpha \!-\! \sigma \!-\! t \right)
     }
     {\Gamma\left( \frac{n}{2}L \!-\! \alpha \!-\! t \right)
     }
\times
\frac{
      \Gamma\left( \alpha \!-\! \frac{n}{2}(L-1) \!+\! t \right)
     }
     {
      \Gamma\left( \alpha \!+\! \sigma \!-\! \frac{n}{2}(L-1) \!+\! t \right)
     } \;. 
$$
Let us introduce two differential operators, $K_1$ and $K_2$, defined as 
$$
K_1:
\prod_{j=0}^{\sigma-1} [(\sum_{k=1}^L \theta_k + b + j)]  \;, 
\quad 
K_2:
\prod_{j=1}^{\sigma} [(-\sum_{k=1}^L \theta_k + a - j)] \;,
$$
where 
$a = \frac{n}{2}L-\alpha$ and  $b=\alpha-\frac{n}{2}(L-1)$. 
Applying these operators to Eq.~(\ref{mellin-barnes:v}), we get 
$$
\left( K_1 \circ K_2 \right) V^{(L)} = J^{(L)},  
$$
where $J^{(L)}$ is the sunrise diagram defined by
Eq.~(\ref{mellin-barnes:sunset}). 
Indeed, we have 
$$
K_1 \circ K_2  
\left( 
\prod_{j=1}^L z_j^{t_j} 
\left( 
\frac{\Gamma\left(b \!+\! t \right)}
     {\Gamma\left(b \!+\! \sigma \!+\! t \right)          }
\right) 
\left( 
\frac{\Gamma\left(a \!-\! \sigma  \!-\! t \right)}
     {\Gamma\left(a \!-\! t \right)          }
\right) 
\right) 
= \prod_{j=1}^L z_j^{t_j} \;.
$$
The operators $K_1$ and $K_2$ are products of differential operators of the first order, 
so that their inverse operators correspond to one-fold integrals over linear forms.
Integrals of this type convert Puiseux-type solutions of diagram $J^{(L)}$ into
Puiseux-type solutions of diagram $V^{(L)}$.
As a consequence, the dimension of the space of nontrivial solutions of the
differential operators related to the $V$-type diagram
coincides with the dimension of the space of 
nontrivial solutions of the differential operators related to
the sunrise diagram. 
$\square$

\section{Sunrise diagram and Bessel functions} 
\label{appendixB} 
Let us consider the one-fold integral representation of the sunrise diagram 
according to Ref.~\cite{Mendels} (see also Ref.~\cite{sunrise:2loop}). 
Using the Fourier transform of the massive propagator in Euclidean space-time,  
\begin{eqnarray}
\Delta(x,M) & \equiv & \int d^n k \frac{\exp{(-ikx)}}{(k^2+M^2)^\alpha}
= 
\frac{ 2 \pi^{\tfrac{n}{2}}}{\Gamma(\alpha)} 
\left( \frac{x}{2} \right)^{\alpha-\tfrac{n}{2}} 
     M^{\tfrac{n}{2}-\alpha} K_{\tfrac{n}{2}-\alpha}(M x)
\nonumber \\ & \sim & 
\left( \frac{M}{x} \right)^{\frac{n}{2}-\alpha} K_{\tfrac{n}{2}-\alpha}(M x) \;,
\end{eqnarray}
where $K_\nu(z)$ is the MacDonald function (for details, see Section~$3.7$ in
Ref.~\cite{watson})
and $n$ is non-integer, and performing the angular integration in $n$
dimensions, 
$$
\int d^n \hat{x} \exp{(-ikx)} = 2 \pi^{\tfrac{n}{2}} 
\left( \frac{x \sqrt{-p^2}}{2} \right)^{1-\frac{n}{2}} 
J_{\tfrac{n}{2}-1}(x \sqrt{-p^2}) \;,  
$$
where  $J_\nu(z)$ is the Bessel function,
we find the one-fold integral representation of the massive sunrise diagram
in Euclidean space-time to be
\begin{eqnarray}
J^E(\vec{M_j^2};\vec{\alpha_j};p^2)
& = &  
\int 
\prod_{j=1}^L \frac{d^n k_j}{[k_j^2\!+\!M_j^2]^{\alpha_j} } 
\times
\frac{1}{[(p\!-\!k_1\!-\cdots\!-\!k_L)^2\!+\!M_{L+1}^2]^{\alpha_{L+1}}} \;, 
\\ 
& \sim & 
\int_0^\infty \frac{dt}{t^{\tfrac{n}{2}L - \sum_{k=1}^{L+1} \alpha_k}} 
\times              J_{\tfrac{n}{2}-1}(t \sqrt{-p^2} )  
\times  \prod_{j=1}^{L+1} K_{\tfrac{n}{2}-\alpha_j}(M_j t) 
\;. 
\nonumber 
\label{sunrise:bessel}
\end{eqnarray}
For the massless propagator, we have
\begin{eqnarray}
\lim_{z \to 0} K_\nu(z) & \to  & \frac{1}{2} \left(\frac{z}{2}\right)^{-\nu} \Gamma(\nu) + \it{O}(z^{-\nu+1}) \;,
\nonumber \\ 
\Delta(x,0) & \equiv & \int d^n k \frac{\exp{(-ikx)}}{(k^2)^\alpha}
\sim \left(\frac{1}{x^2} \right)^{\frac{n}{2}-\alpha} \;,
\nonumber 
\end{eqnarray}
so that 
\begin{eqnarray}
J_R^E(\vec{M_j^2};\vec{\alpha_j},\vec{\beta_i};p^2)
& = &  
\int \prod_{j=1}^R \frac{d^n (k_1 \cdots k_L)}
{
[k_j^2\!+\!M_k^2]^{\alpha_j} 
[k_{R+1}^2]^{\beta_1} 
\cdots
[(p\!-\!k_1\!-\cdots\!-\!k_L)^2]^{\beta_{L+1-R}}   
} 
\nonumber \\ 
& \sim &  
\int_0^\infty \frac{dt}{t^{\tfrac{d}{2}L \!-\! \sum_{k=1}^{R} \alpha_k \!-\! \sum_{j=1}^{L+1-R} \beta_j }} 
              \times
              J_{\tfrac{n}{2}-1}(t \sqrt{-p^2} )  
              \times              
              \prod_{j=1}^{R} K_{\tfrac{n}{2}-\alpha_j}(M_j t) 
\;. 
\nonumber \\ && 
\label{sunrise:bessel:R}
\end{eqnarray}
The recurrence relations for the MacDonald functions,
$$
K_{\nu-1}(z) = K_{\nu+1}(z) - \frac{2}{z} \nu K_\nu(z)\;, 
\quad
\frac{d}{dz} K_{\nu}(z) = -\frac{1}{2} \left( K_{\nu-1}(z) + K_{\nu+1}(z) \right)\;, 
$$
in combination with recurrence relations for the Bessel functions, 
$$
J_{\nu-1}(z) = -J_{\nu+1}(z) + \frac{2}{z} \nu J_\nu(z)\;, 
\quad  
\frac{d}{dz} J_{\nu}(z) = \frac{1}{2} \left( J_{\nu-1}(z) - J_{\nu+1}(z) \right)\;, 
$$
allow us to change the power of $t$ and the orders of the Bessel and MacDonald functions to any integer values. 
The result of Section~\ref{dimension} can be applied to evaluate the 
dimension of the basis within this reduction.
If all the values of $\alpha_j$ and $\beta_k$ are integer, then the results of
Section~\ref{dimension} are valid. 
In particular, the difference between the reducible and irreducible monodromies 
of the integrals defined by Eqs.~(\ref{sunrise:bessel}) and (\ref{sunrise:bessel:R}), respectively,
is an extra integer power of the variable $t$.
There are a few other cases where the results of Section~\ref{dimension} are
applicable, namely, 
different combinations of integer and non-integer values of $\alpha_i$ and
$\beta_j$.
In the following, we present a few examples.
\begin{enumerate}
\item
Under the conditions that 
\begin{itemize}
\item  
$\tfrac{n}{2} \notin \mathbb{Z}$ \;,  \quad 
\item  
$\alpha_j \notin \mathbb{Z}$, \quad for $\forall j=1, \cdots, L+1$ \;, 
\item  
$\tfrac{n}{2} - \alpha_a \notin \mathbb{Z}$,  \quad for $\forall a=1, \cdots, L+1$ \;, 
\item
$\sum_{S} \alpha_k \notin Z$, for any subset of $\forall k=1, \cdots, R$, \\
\end{itemize}
there is an invariant subspace of dimension 1 for 
the integral defined by Eq.~(\ref{sunrise:bessel}).  
\\
Indeed, in this case, only one equation, 
$
\frac{n}{2}(L+1) - \sum_{S_2} \frac{n}{2} = 0 \quad \left(\mbox{\rm mod} ~~ \mathbb{Z} \right)
$,
is valid. 
In this case, the number of irreducible integrals defined by Eq.~(\ref{sunrise:bessel})
is equal to $2^{L+1}-1$.

\item
Under the conditions that 
\begin{itemize}
\item  
$n \in \mathbb{Z}$,  \quad 
\item  
$\beta_a \in \mathbb{Z}$,  $\forall a=1, \cdots, L+1-R$ , \quad 
\item  
$\alpha_j \notin \mathbb{Z}$, \quad for $\forall j=1, \cdots,R$
\item
$\sum_{S} \alpha_k \notin Z$, for any subset of $\forall k=1, \cdots,R$, \\
\end{itemize}
there is an invariant subspace of dimension 1 for the integral defined by
Eq.~(\ref{sunrise:bessel:R}). 
\end{enumerate}

\acknowledgments
We are grateful to V.V.~Bytev for discussions on similar problems related with
differential reduction to and R.N.~Lee for cross-checking some of our results.
We thank A.I.~Davydychev for useful discussions and for carefully reading the
manuscript.
We are grateful to the anonymous Referee for his very constructive comments. 
This work was supported in part by the
German Research Foundation DFG through the Collaborative Research
Centre No.~676 {\it Particles, Strings and the Early Universe---The
Structure of Matter and Space-Time}.
The work of MYK was supported in part by the Heisenberg-Landau Program (Dubna, Russia).

%
%



\end{document}